# $IrSr_2Sm_{1.15}Ce_{0.85}Cu_{2.175}O_{10}$: A Novel Reentrant Spin-Glass Material

R. H. Colman and A. C. Mclaughlin*

*Department of Chemistry, University of Aberdeen, Meston Walk, Aberdeen AB24 3UE, UK*

*a.c.mclaughlin@abdn.ac.uk

**Abstract**

A new iridium containing layered cuprate material, $Ir_{0.825}Sr_2Sm_{1.15}Ce_{0.85}Cu_{2.175}O_{10}$, has been synthesized by conventional ambient-pressure solid-state techniques. The material's structure has been fully characterized by Rietveld refinement of high resolution synchrotron X-ray diffraction data; tilts and rotations of the $IrO_6$ octahedra are observed as a result of a bond mismatch between in-plane Ir-O and Cu-O bond lengths. DC-susceptibility measurements evidence a complex set of magnetic transitions upon cooling that are characteristic of a reentrant spin-glass ground-state. The glassy character of the lowest temperature, $T_g$=10 K, transition is further confirmed by AC-susceptibility measurements, showing a characteristic frequency dependence that can be well fitted by the Vogel-Fulcher law and yields a value of $\frac{\Delta_{T_f}}{[T_f \Delta \log(\omega)]} = 0.015(1)$, typical of dilute magnetic systems. Electronic transport measurements show the material to be semiconducting at all temperatures with no transition to a superconducting state. Negative magnetoresistance is observed when the material is cooled below 25 K, and the magnitude of this magnetoresistance is seen to increase upon cooling to a value of MR = -9 % at 8 K.

# I. INTRODUCTION

The layered cuprates have attracted much scientific interest in recent years. The 1212- and 1222- layered cuprates have been well studied and have the general formulas $MA_2RECu_2O_{8-\delta}$ and $MA_2RE_2Cu_2O_{10-\delta}$ respectively, where $M$ is commonly a transition or p-block metal, $A$ is an alkaline earth ion and $RE$ is a rare earth or mixture of rare earth ions. The δ signifies the oxygen non-stoichiometry which is dependent on synthesis conditions, $M$ oxidation state and the Ln:Ce ratio,[1] a key tool for controlling the oxidation state of the copper planes and inducing superconductivity. The flexibility of this structural motif has lead to an explosion of new materials, including analogues incorporating $M$-site ions varying from: first row transition metals such as Cu, Fe and Co;[2,3,4] second row transition metals such as Ru;[5] third row transition metals such as Ta, Hg as well as p-block elements such as Ga and Tl.[6,7]

This wide variation in chemical composition leads to an equally wide diversity of electronic properties. As well as superconductivity, the fine electronic balance in many of these materials leads to novel and unexpected phenomena. The ruthenocuprates have received much attention for the observation of simultaneous weak ferromagnetism and superconductivity, two properties thought incompatible.[8] When underdoped, the ruthenocuprates have also been shown to display a sizeable low temperature magnetoresistance.[9,10,11,12] Surprisingly this magnetoresistance is found to be tunable from negative to positive values as $<r_A>$, the mean $A$ site ($R_{1.1}Ce_{0.9}$) cation radius, decreases in a series of $RuSr_2R_{1.1}Ce_{0.9}Cu_2O_{10-\delta}$ ($R$ = Nd, Sm, Eu, and Gd with Y) samples.[13] This lattice effect is further evidenced from studies of $Ru_{1-x}Ta_xSr_2Nd_{0.95}Y_{0.15}Ce_{0.9}Cu_2O_{10}$ materials; $MR_{9T}$(4 K) increases from -28% to -49% as $x$ increases from 0 – 0.2 which further expands the unit cell.[12]

In an attempt to broaden this library of layered cuprates and search for further novel properties, we have experimented with the substitution of alternative transition metals into the 1222 layered structure. The large spin-orbit interaction in the heavy transition metals has been recently proposed to lead to exotic physics in a number of iridate systems. These exotic physics range from topological Fermi-arc states in $Sr_2IrO_4$[14] to long range order in $Na_2IrO_3$[15] and even a proposed 3-dimensional quantum spin liquid ground-state in the frustrated hyperkagome lattice compound $Na_4Ir_3O_8$.[16] The possibility of such unusual physical phenomena make iridium-containing cuprates an obvious synthetic target.

Synthesis of several 1212-type iridocuprates by high pressure have previously been reported,[17,18,19] but this approach limits both sample volume and ease of compositional variation so that superconductivity has not yet been observed in the 1212-irridocuprates. The 1222-type iridocuprates have not previously been synthesized. Whilst $Ir^{5+}$ containing double perovskites, having the electronic configuration $[Xe]4f^{14}5d^4$ (where [Xe] is the xenon core), have found no contribution of the Ir ion to the magnetic properties,[20] the $J=1/2$ $Ir^{4+}$ ion ($[Xe]4f^{14}5d^5$) has been seen to magnetically order in related double-perovskite materials.[21]

In this paper we report the ambient-pressure synthesis, structural characterisation and properties of the novel 1222-type iridocuprate, $Ir_{0.825}Sr_2Sm_{1.15}Ce_{0.85}Cu_{2.175}O_{10}$. Whilst the sample is not found to be superconducting under these synthetic conditions a complex temperature dependence to the magnetisation, studied using both AC- and DC-susceptibility measurements, suggests reentrant spin glass behavior, likely due to disordered occupation of $Cu^{3+}$, $Ir^{4+}$ and $Ir^{5+}$ ions across the nominally $IrO_2$ layer. Low temperature negative magnetoresistance is also observed upon cooling below 25 K.

## II. EXPERIMENTAL

The sample of $Ir_{0.825}Sr_2Sm_{1.15}Ce_{0.85}Cu_{2.175}O_{10}$ was prepared by standard solid state techniques. A stoichiometric mixture of $IrO_2$, $SrCO_3$, $Sm_2O_3$, $CeO_2$, and CuO powders were combined by thorough grinding in an agate mortar. Prior to use the $Sm_2O_3$ and $CeO_2$ reagents were calcined at 800°C for 12 hours to remove any absorbed moisture and $CO_2$. The mixture was pressed into a 10 mm pellet, loaded into a ceramic alumina crucible and placed in a muffle furnace before heating to 800°C for 10 hours. The pellet was then reground, repelleted and transferred to a tube furnace. The sample was then heated under flowing oxygen to 1020°C at a rate of 5°C/minute, allowed to dwell for 14 hours and cooled at the same rate before regrinding and repeating this oxygen firing a further two times. Care was taken with furnace calibration and sample positioning as the synthesis was found to be extremely sensitive to the firing temperature. As little as ±10°C was found to lead to dramatically increased impurity formation. Room temperature laboratory X-ray diffraction patterns were collected on a Bruker D8 Advance diffractometer. Data were collected over the range $5° < 2\theta < 100°$, with a step size of $0.02°$. The profiles could all be indexed on a tetragonal *I4/mmm* symmetry space group as previously reported for other members of the 1222 layered cuprate family.[1,3]

## III. RESULTS AND DISCUSSION

### A. Synchrotron X-ray diffraction

To confirm phase purity and refine the crystal structure, high resolution synchrotron X-ray diffraction data was collected on a sample of $Ir_{0.825}Sr_2Sm_{1.15}Ce_{0.85}Cu_{2.175}O_{10}$. A pattern was collected at 290 K on beamline ID31, ESRF. A wavelength of $\lambda = 0.40002$ Å was used, and the sample was contained in a 0.5 mm-diam borosilicate glass capillary mounted on the axis of the diffractometer, about which it was spun at ~1 Hz to improve the powder averaging of the crystallites. Diffraction patterns were collected over the angular range $2° < 2\theta < 50°$ and rebinned to a constant step size of $0.002°$ for each scan. The high-angle regions of the pattern were scanned several times to improve the statistical quality of the data. The Rietveld refinement program GSAS was used to refine the structure.[22] The data and refinement fits can be seen in Fig. 1(a). This led to the refined lattice parameters and atomic positions seen in Table I.

Whilst not detectable by laboratory X-ray diffraction, the higher resolution and increased flux of the synchrotron data showed the presence of impurity peaks that could be confidently assigned to the minor impurity phases $Sr_3IrCuO_6$, $Sr_2CeIrO_6$ and unreacted $Sm_2O_3$. The small volume-fractions of these impurities are not expected to effect the bulk property measurements and so do not change the conclusions of the measurements discussed later in this paper as these materials do not exhibit magnetic transitions coincident with the transitions reported. The structure can be described as alternating layers of nominally $IrO_2$, SrO, $CuO_2$, Sm/Ce, $O_2$, Sm/Ce, $CuO_2$, SrO and $IrO_2$ respectively (Fig. 1(b)). Results from Rietveld refinement of the synchrotron X-ray diffraction data are displayed in Table 1 and Supplementary Table 1 [23]. As with all compounds of the 1222 structure-type, occupation of the rare-earth site by Sm and Ce ions is found to be fully disordered.[1,3]

Attempts to refine the Ir occupation of the 2$a$ Wyckoff site lead to a preference for reduced occupation. Site vacancies are unlikely, due to electrostatic arguments, leaving the possibility of mixed occupation of the site by both Ir and a weaker scatterer. Consideration of the available cations leaves Cu as the most likely candidate, primarily due to similarities in ionic radii (0.54 Å and 0.57 Å for $Cu^{3+}$ and $Ir^{5+}$ respectively) [24] and preferred octahedral coordination environments. Refinement of mixed Ir/Cu occupation leads to a substitution of 17.5(5)% Cu onto the nominally Ir site. There is no evidence of cation disorder within the $CuO_2$ planes.

Particular attention was paid to the refinement of oxygen positions as these define the octahedral rotation and tilting. Whilst ordered tilting of nominally-$IrO_6$ octahedra is not symmetry allowed within the *I4/mmm* space group, refinements of the apical and equitorial oxygens of the octahedra, O(1) and O(3) respectively, found unphysical thermal displacement parameters when placed on the high symmetry 4$e$ and 4$c$ Wyckoff sites. There was no evidence of weak superstructural peaks which would indicate a change in symmetry due to ordered tilting/rotations of the $IrO_6$ octahedra. An improved fit and physically reasonable thermal displacement parameters was found by splitting the sites as shown in Table 1. The refined positions of the split oxygen sites indicates both a disordered rotation and tilting of the $IrO_6$ octahedra of 14.6(3)° and 7.6(6)°, respectively. Similar tilts and rotations of the $RuO_6$ octahedra are observed in $RuSr_2GdCu_2O_8$ [5] and $RuSr_2Gd_{2-x}Ce_xCu_2O_{10-\delta}$.[1]

Attempts to refine the occupation of these oxygen sites resulted in stable but non-unique solutions, likely due to correlations with both site positions and thermal parameters. As a result, the O(1) and O(3) positions were fixed to complete occupancy. Additionally oxygen site O(4), residing within the rare-earth block, refined to complete occupancy and so was fixed at unity. This oxygen position is most commonly found to accommodate oxygen vacancies in the 1222-type structures,[1, 25] but due to the relatively high Ce content in this material vacancies are not required for charge balancing.

Bond valence analysis, using the program VaList,[26] of the Cu (*4e*) site results in an average bond valence parameter of 2.45. Whilst this value is larger than would be expected for typical $Cu^{2+}$ coordination it is typical of layered ruthenocuprate materials[1] and cannot be taken as evidence for increased hole doping of the $CuO_2$ plane. Consideration of the Cu-O bond distances 2.155(11) Å and 1.9368(7) Å for Cu-O(1) and Cu-O(2), respectively, find them to be comparable to those of the 1222-ruthenocuprates, where for example values of 2.157(5) Å and 1.9236(3) Å, respectively, are reported in the closely related compound $RuSr_2Gd_{1.3}Ce_{0.7}Cu_2O_{10}$.[1] Due to the mixed occupation as well as possible mixed valency of the nominally Ir (*2a*) site, bond valence analysis cannot be confidently performed. The Ir-O bond lengths give values of 1.972(11) Å and 1.992(3) Å for Ir-O(1) and Ir-O(3), respectively. These values are again similar to those of the ruthenocuprates, unsurprisingly due to the similar ionic radii of the two ions (0.57 Å and 0.565 Å for octahedral $Ir^{5+}$ and $Ru^{5+}$ respectively),[24] leading to comparable octahedral coordination environments. The observed rotations and tilts of the $IrO_6$ octahedra therefore arise due to the bond mismatch between in plane Ir-O(2) and Cu-O(2).

### B. DC-SQUID magnetometry

To probe the bulk magnetic properties of this new material, measurements of magnetization vs. both field and temperature were performed using a Quantum Design MPMS-XL magnetometer. Measurements against temperature were performed in the range $2 < T < 300$ K after cooling both in zero-field (ZFC) and in field (FC) using an applied field of $H_{DC}$=100 Oe. Isothermal measurements against field were made across the range $-7 < H < 7$ T at temperatures of T = 2, 60 and 200 K.

Two distinct magnetic transitions are evidenced by the $\chi$ vs. *T* measurements (Fig. 2): a broad feature, seen in the inset of Fig. 2 and labeled as $t_2$, centered at 120 K that shows clear divergence between ZFC and FC datasets; and a sharp increase in susceptibility upon cooling through 8 K, labeled as $t_3$, with further ZFC-FC bifurcation. Inspection of a plot of $\chi^{-1}$ vs. *T* (Fig. 2(b)) shows that ZFC and FC splitting is also observed above $t_2$ and is persistent up to 300 K. This splitting evidences history dependence to the magnetization that can be explained by the presence of ferromagnetic clusters in an otherwise paramagnetic state.

This history dependence is further confirmed by the *H* vs. *T* measurements seen in Fig. 3. At 200 K a very weak hysteresis is observed but dominated by the linear paramagnetic contribution. Below $t_2$, at 60 K, the hysteresis loop increases in size, and below $t_3$ becomes much more evident. At 2 K the loop takes an unconventional shape known as a wasp-waist that has previously been assigned to disorder in the magnetic exchange.[27]

This complex type of magnetic behavior is commonly observed in reentrant spin-glasses (RSGs).[28] Upon cooling a reentrant spin-glass there is an initial onset of partial order at $t_1$ (>300 K in $Ir_{0.825}Sr_2Sm_{1.15}Ce_{0.85}Cu_{2.175}O_{10}$) where ferromagnetic clusters are formed in the otherwise paramagnetic state. The size of these clusters grows upon cooling until an ordering transition is observed at $t_2$. As the material is cooled further a third, disordering, spin-glass transition is seen at $t_3$.

### C. AC-SQUID magnetometry

To further probe the low temperature transition, $t_3$, frequency dependent AC-susceptibility measurements were performed, again using a Quantum Design MPMS-XL magnetometer. A drive field of $H_{ac}$ = 3.5 Oe was used, and measurements were performed at frequencies spanning several decades. The temperature dependence of the real and imaginary components to the susceptibility, χ' and χ'' respectively, can be seen in Fig. 4. A peak is observed in both components, that shows frequency dependence in the peak position as well as magnitude.

The peaks in χ' were fitted with a Gaussian function to obtain the frequency dependence of the transition temperature, $T_f$, taken as the peak maximum. This frequency dependence of $T_f$ can be seen in Fig. 5. For spin-glass transitions $T_f$ is expected to follow the Vogel-Fulcher law, $\omega = \omega_0 e^{[-E_a/k_B(T_f-T_o)]}$, where ω is the AC-frequency, $E_a$ is the activation energy of the spin glass, $\omega_0$ is the characteristic frequency for spin freezing usually in the range $10^8$ - $10^{15}$ and $T_0$ is the Vogel-Fulcher temperature.[29] Fits to the Vogel-Fulcher law gave excellent agreement with the experimental data, as can be seen in Fig. 5, and resulted in the following refined parameters: $\omega_0 = 5.2 \times 10^{15}$ Hz; $\frac{E_a}{k_B} = 65.1(3)$ K and $T_0 = 7.748(8)$ K. The excellent agreement of this fit is strong evidence that $t_3$ is a spin-glass transition, strengthening the hypothesis that the complex set of magnetic transitions seen in $Ir_{0.825}Sr_2Sm_{1.15}Ce_{0.85}Cu_{2.175}O_{10}$ can be assigned to reentrant spin-glass behavior.

Another consideration in the designation of spin glasses is the calculation of $\Delta_{T_f}/[T_f \Delta \log(\omega)]$.[29] In canonical spin glasses this value falls in the range 0.005 - 0.28 with $Ir_{0.825}Sr_2Sm_{1.15}Ce_{0.85}Cu_{2.175}O_{10}$ having a value of 0.015(1), in line with typical dilute magnetic systems such as PdMn and NiMn having 0.013 and 0.018, respectively.[30,31] Competition in magnetic exchange between antiferromagnetic and ferromagnetic interactions is the origin of the RSG phenomenon and this competition is presumably a consequence of the presence of $Ir^{4+}$, $Ir^{5+}$ and $Cu^{3+}$ within the *nominally-*$IrO_2$ plane for $IrSr_2Sm_{1.15}Ce_{0.85}Cu_2O_{10-\delta}$. These competing interactions lead to frustrations between the spins so that below 9.6 K ($T_f$) a breakdown of the higher temperature ferromagnetic state occurs and a disordering spin-glass transition is observed. Furthermore such RSG behavior has not been previously evidenced in the $CuO_2$ plane of the many underoped cuprates reported, so that is seems most likely that the origin of the RSG is in the $IrO_2$ slab.

### D. Electron-transport mesurements

The electronic transport properties of a sintered ceramic bar of $Ir_{0.825}Sr_2Sm_{1.15}Ce_{0.85}Cu_{2.175}O_{10}$ were investigated using a Quantum Design Physical Property Measurement Sytem (PPMS). The resistivity of the bar was measured, using the 4-point probe technique, as a function of temperature in the range $4 < T < 350$ K in both zero applied field and with a 7 T applied field. Additionally an isothermal measurement of the resistivity was made at 8 K, sweeping the field range $0 < H < 9$ T.

The resistivity measurements in zero field found the material to be semiconducting ($\rho_{290K} = 0.019$ $\Omega$.cm) and there was no evidence of superconductivity which suggests that the $CuO_2$ planes are underdoped (i.e. that the oxidation state of Cu < 2.05 [32]). It can be seen in Fig. 6 that between 350 K and 65 K there is an excellent fit to Mott 3-dimensional variable-range-hopping ($\rho = \rho_0 \exp(T_0/T)^{1/4}$, where $\rho$ is the measured resistivity, $T$ is the temperature and $T_0$ is a fitted localization temperature).[33] Upon cooling below 70 K there is a distinct change in slope of the linear fits so that $T_0 = 6.25 \times 10^6$ K and $6.2 \times 10^5$ K in the high and low temperature regions respectively. This indicates a change in electronic localization length upon cooling ($T_0 = \lambda\alpha^3/k_B N(E_F)$ where $T_0$ is the degree of electronic disorder, $\lambda$ is a dimensionless constant, $\alpha^{-1}$ is equal to the localisation length, $k_B$ is the Boltzmann constant and $N(E_F)$ is the density of localised states at $E_F$).

Using measurements of the resistivity measured in both zero field, $\rho_{H=0}$, and in a 7 T field, $\rho_{H=7}$, the temperature dependent magnetoresistance can be calculated. Magnetoresistance is commonly defined as the percentage change in resistance upon application of a field, $MR_{H=7T} = 100 \times (\frac{\rho_{H=7} - \rho_{H=0}}{\rho_{H=0}})$. At high temperatures, no discernable magnetoresistive effects are observed. Upon cooling below 20 K a negative magnetoresistance is observed which increases in magnitude upon cooling (Fig. 7). The isothermal field sweep at 8 K, shown in the inset of Fig. 7, confirms the magnetoresistive effect and gives a maximum value of $MR = -8$ % at $H = 9$ T. It would appear that low temperature negative magnetoresistance is a common feature of the metallocuprates ($M$Sr$_2$Ln$_{2-x}$Ce$_x$Cu$_2$O$_{10-\delta}$; $M$ = Ir, Ru,[11-14] Co[34] and is a result of an increase in magnetopolaron mobility upon application of a magnetic field.[10, 12, 13] Finally we note that these results demonstrate that the RSG behavior in the IrO$_2$ planes are less effective at enhancing this magnetopolaron mobility than the weak ferromagnetism in the RuO$_2$ plane of RuSr$_2$Ln$_{1-x}$Ce$_x$Cu$_2$O$_{10-\delta}$ where magnetoresistances of up to -49% have been reported [9, 12, 13].

## IV. CONCLUSIONS

In summary we report the synthesis of the new 1222-iridocuprate material Ir$_{0.825}$Sr$_2$Sm$_{1.15}$Ce$_{0.85}$Cu$_{2.175}$O$_{10}$. High resolution structural studies using synchrotron X-ray diffraction show the material to crystallise in the space-group *I4/mmm*, common to the 1222-cuprates. The structural refinement finds there to be a disordered tilting and rotation of the *nominally*-Ir octahedra as well as partial Cu substitution onto the Ir site. Magnetic property measurements of both AC and DC susceptibility show a complex temperature dependence that is characteristic of a reentrant spin glass ground-state. This behaviour is likely due to the disordered occupation of magnetically active $Cu^{3+}$ and $Ir^{4+}$ in an otherwise diamagnetic lattice of $Ir^{5+}$ on the Ir site, leading to frustrated exchange interactions. Electronic transport measurements find the material to be semiconducting at all measured temperatures, suggesting the $CuO_2$ planes are underdoped, and resistivity measurements in an applied field find there to be an increase in the magnitude of negative magnetoresistance upon cooling below 20 K. The features observed in the magnetic property and electronic transport measurements are not obviously concomitant, highlighting the distinctly different origins of the phenomena as coming from the $IrO_2$ and $CuO_2$ planes, respectively. Future work will concentrate on the synthesis of superconducting Ir-1222 material as this new iridocuprate would be an ideal material for further study of the interaction between discreet (ferro)magnetic and superconducting planes.

## Acknowledgements

We acknowledge EPSRC for provision of research grant number EP/F035225; STFC-GB for provision of beam-time at the ESRF and A. Hill for assistance with the synchrotron X-ray diffraction experiments.

TABLE I. Refined structural parameters for Ir$_{0.825}$Sr$_2$Sm$_{1.15}$Ce$_{0.85}$Cu$_{2.175}$O$_{10}$ in the space-group *I4/mmm* from Rietveld refinement of synchrotron X-ray diffraction data.

| a (Å) | c (Å) | Volume (Å$^3$) | $R_{wp}$ | $R_p$ | $\chi^2$ | | | |
|---|---|---|---|---|---|---|---|---|
| 3.85647(2) | 28.2347(2) | 424.378(4) | 0.1009 | 0.077 | 1.849 | | | |

| Atom(s) | Wyckoff site | x | y | z | Occupancy | $U_{iso}$ (Å$^2$) | $U_{11}$ (Å$^2$) [a] | $U_{33}$ (Å$^2$) [a] |
|---|---|---|---|---|---|---|---|---|
| Ir , Cu | 2a | 0 | 0 | 0 | 0.825(5) , 0.175(5) | - | 0.0030(3) | 0.0037(4) |
| Sr | 4e | 0.5 | 0.5 | 0.07839(5) | 1 | - | 0.0103(3) | 0.0135(6) |
| Sm , Ce | 4e | 0.5 | 0.5 | 0.20486(2) | 0.575 , 0.425 | - | 0.0049(3) | 0.0050(3) |
| Cu | 4e | 0 | 0 | 0.14347(7) | 1 | - | 0.0044(3) | 0.0082(7) |
| O(1) | 16m | 0.952(3) | 0.952(3) | 0.0685(4) | 0.25 | 0.014(4) | - | - |
| O(2) | 8g | 0 | 0.5 | 0.1499(3) | 1 | 0.010(1) | - | - |
| O(3) | 8j | 0.870(3) | 0.5 | 0 | 0.5 | 0.015(3) | - | - |
| O(4) | 4d | 0 | 0.5 | 0.25 | 1 | 0.011(2) | - | - |

[a] All of the anisotropically refined sites have $U_{11}=U_{22}\neq U_{33}$, $U_{12}=U_{13}=U_{23}=0$

**Figure Captions**

FIG 1 (a). Synchrotron X-ray diffraction pattern and Rietveld refinement fit to data collected on $Ir_{0.825}Sr_2Sm_{1.15}Ce_{0.85}Cu_{2.175}O_{10}$. The crosses show the collected pattern, the upper line is the refined fit, the lower line is the difference and the tick marks denote reflection positions for $Sr_3IrCuO_6$, $Sr_2CeIrO_6$, $Sm_2O_3$ and $Ir_{0.825}Sr_2Sm_{1.15}Ce_{0.85}Cu_{2.175}O_{10}$, from top to bottom respectively. The final goodness of fit parameter, $\chi^2 = 1.849$. The inset shows a blow-up of the high-angle region, again displaying an excellent fit to the data. (b). The crystal structure of nominally $IrSr_2(Sm,Ce)_2Cu_2O_{10}$ showing alternating layers of $Sr_2IrO_6$, $(Sm,Ce)_2O_2$ and $CuO_2$ common to all 1222-type materials.

FIG 2 (a). A plot of susceptibility *vs.* temperature measurements showing both zero-field-cooled (black closed circles) and field-cooled (red open circles) datasets. Transitions are evident at 125 and 10 K. Inset is an expansion highlighting the transition evident at 125 K. (b). A plot of inverse susceptibility *vs.* temperature, clearly showing the ZFC-FC separation which is persistent up to 300 K.

FIG 3. A plots of isothermal hysteresis loops. The 2 K loop shows an unconventional *wasp-waist* shape usually assigned to exchange disorder and the inset is a blow-up of the low field region showing a small hysteresis is still present up to 200 K.

FIG 4. A plot of the real, $\chi'$, and imaginary, $\chi''$, components of the AC-susceptibility across the low temperature transition, $t_3$. At higher frequencies the peak is seen to diminish in magnitude and shift to higher temperatures.

FIG 5. A plot of the fitted peak position, $T_f$, vs. the natural log of the AC-measurement frequency, $\omega$. The solid line shows a fit to the Vogel-Fulcher law, in excellent agreement with the data, indicative of spin-glass behavior.

FIG 6. A plot of the resistivity, $\rho$, vs. $T^{-1/4}$. Linear behavior is indicative of 3-dimensional variable-range hopping. A change in slope indicates a change in the electronic localization length, seen upon cooling below 80 K.

FIG 7. A plot of the *MR* variation with temperature ($H = 7$T) showing an increase in negative magnetoresistance upon cooling below 25 K. The inset shows MR variation with field at 8 K.

**Figures**

FIG. 1.

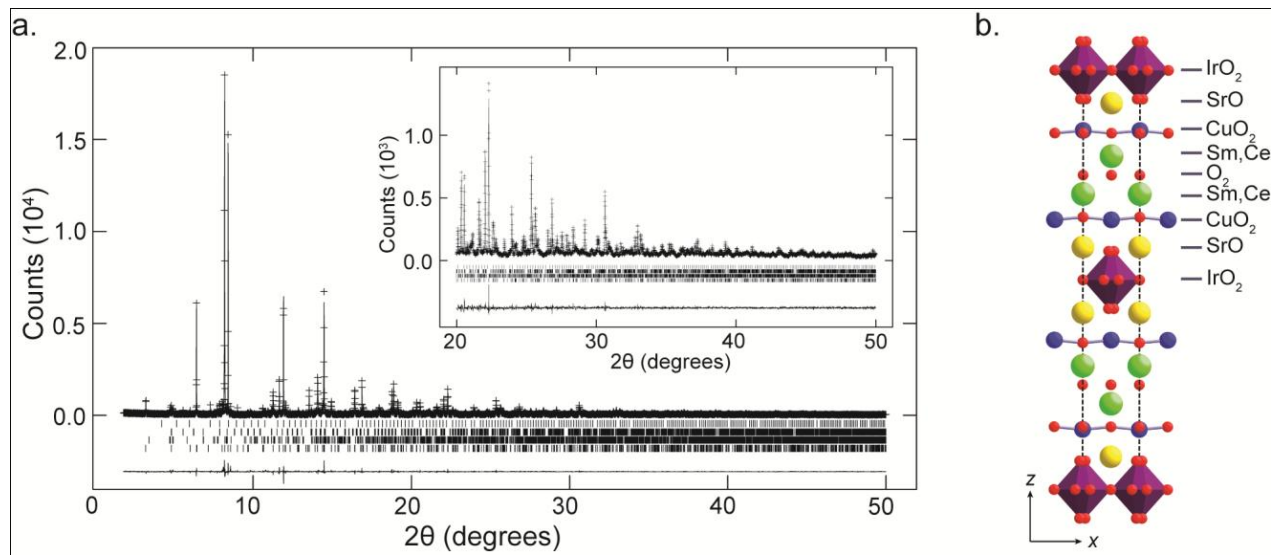

FIG. 2.

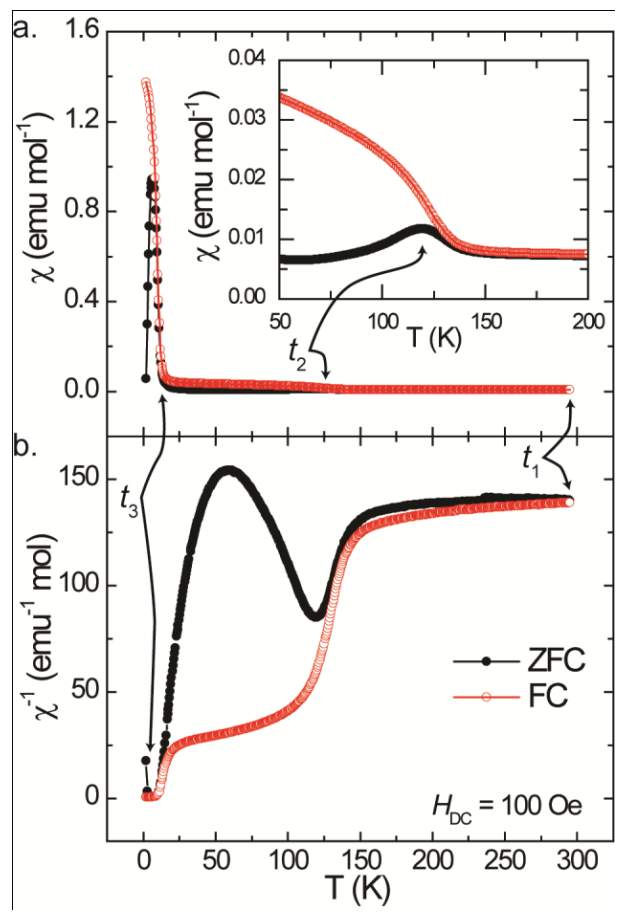

FIG. 3.

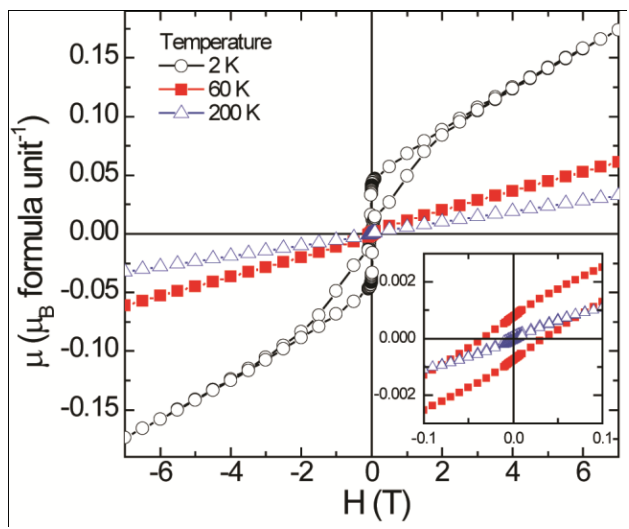

FIG. 4.

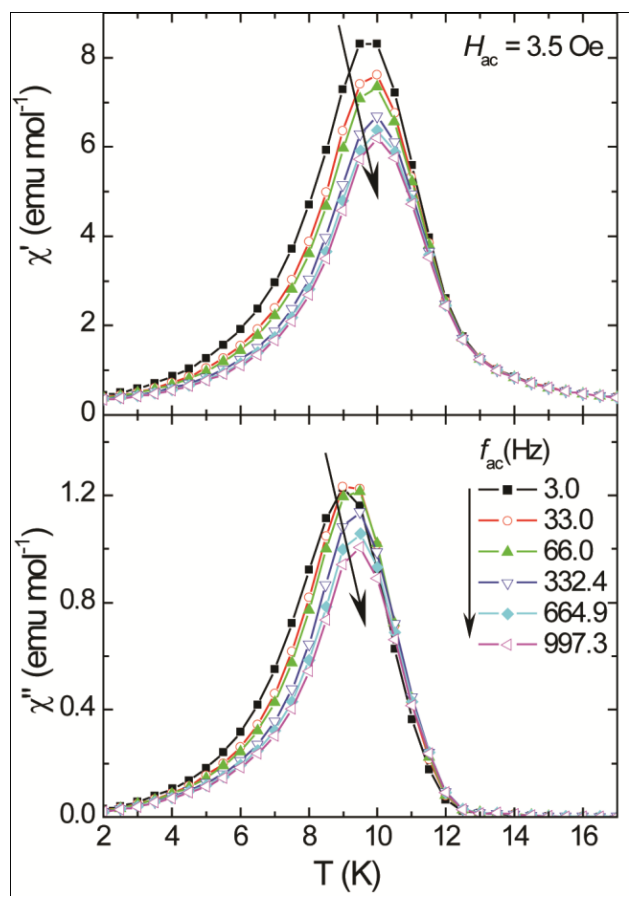

FIG. 5.

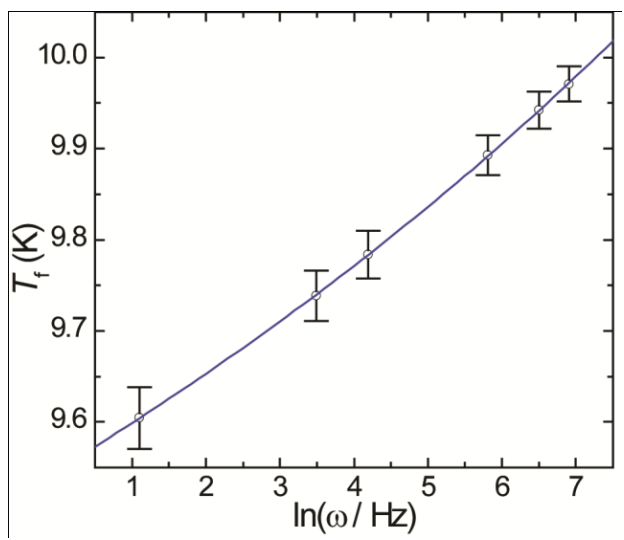

FIG. 6.

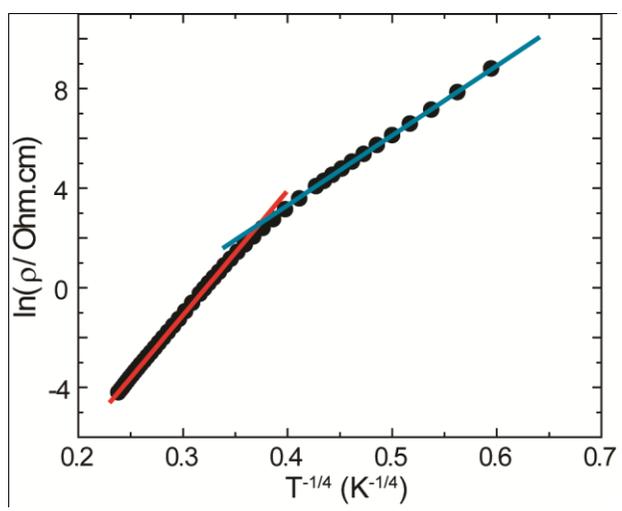

FIG. 7.

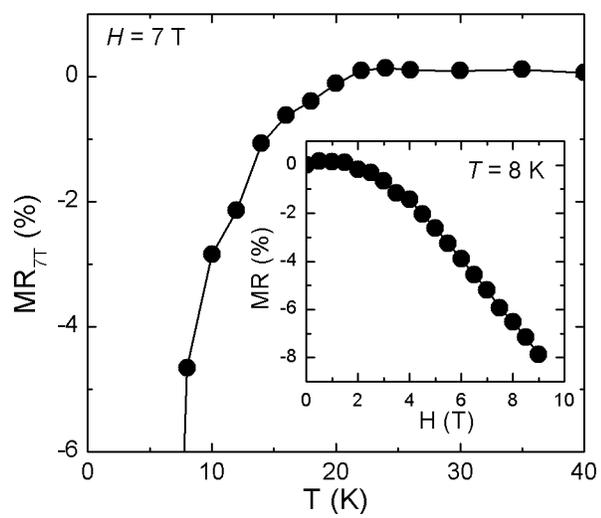